\documentclass{article}

\def\etadg{\eta^\dagger}   
\def\rhodg{\rho^\dagger}
\def\chidg{\chi^\dagger}

\def\chid{\chi^\dagger \chi}
\def\etad{\eta^\dagger \eta}
\def\rhod{\rho^\dagger \rho}
\def\sds{S^\dagger S}

\def\rt2{\frac{1}{\sqrt{2}}}

\begin{document}
\hfill{UM-P-2000/047}
\begin{center}

{\LARGE \bf Generating Neutrino Mass in the 331 Model\\}
\vspace{10mm}
{\bf M.B.Tully 
\footnote{E-mail: mbt@physics.unimelb.edu.au}
and G.C.Joshi
\footnote{E-mail: joshi@bradman.ph.unimelb.edu.au}
}\\
\vspace{10mm}
{\sl Research Centre for High Energy Physics,\\
School of Physics, 
University of Melbourne,\\
Victoria 3010, Australia.}
\end{center}
\vspace{30mm}
\begin{abstract}
A mechanism for generating small tree-level Majorana mass for
neutrinos is implemented in the 331 Model. No additional fermions
or scalars need to be added, and no mass scale greater than a
few TeV is invoked.

\end{abstract}
\newpage 
\section{Introduction}
The recent atmospheric neutrino results from 
Super Kamiokande 
\cite{SuperK}
have
provided the 
long-awaited first
direct evidence for physics beyond the Standard Model.
The conventional interpretation of the
results as due to neutrino oscillation requires
neutrinos to have a very small, but non-zero mass
of around 0.1 eV.

It is a challenge for theories to be able to 
explain why the neutrino mass should be so
much smaller than that of all the other fermions,
and a number of general mechanisms have been
proposed. \cite{Ma1}

However, these methods generally require either
the ad-hoc introduction of additional scalars, as in
the Zee model \cite{Zee}
or other radiative mechanisms, or
the presence of a very large mass scale, as in
the seesaw mechanism 
\cite{Seesaw} or the heavy Higgs triplet
model. \cite{Ma2}
While the latter two solutions are
undoubtedly simple and very elegant, both require
a new mass scale of some $10^{12}$ Gev - the
mass of the right-handed neutrino in the
seesaw model or the mass of the heavy Higgs
in the triplet model. 
Apart from the fact that such a mass scale,
being so widely separated from the electroweak
breaking scale, introduces possible hierarchy
problems, it is also displeasing in that 
it is hard to envisage how physics at such
a high energy scale may be tested in the
forseeable future. \cite{Ma3}

Here we show that a method related to that of 
the Higgs triplet can  be implemented 
with a minimum of contrivance
within the popular 331 Model. \cite{Valle}-\cite{other331}
No heavy mass scale needs to be introduced,
and in fact it is a feature of this model
that in its minimal variant the symmetry
breaking scale is actually constrained to be no
more than a few TeV.
The smallness of the neutrino mass is due
to the small size of lepton-number violating
terms in the scalar potential. 
We emphasise that no additional scalars 
or fermions 
need to be introduced.

\section{The 331 Model}
\subsection{Fermions}
The 331 Model is so-named because the
Standard Model gauge group is extended to
$SU(3)_c \times SU(3)_L \times U(1)_X$.
The various versions 
which 
have been proposed can be characterised in part
by the
choice of the parameter $\xi$ which describes
the embedding of electric charge, 
according to
\begin{equation}
Q=\frac{\lambda_3}{2} + \xi \frac{\lambda_8}{2} +X
\end{equation}
Here, we take $\xi=-\sqrt{3}$.
The leptons of each family then
transform in a triplet with 
$X=0$, namely
\begin{equation}
f_{1,2,3L}=\left( \begin{array}{c}
\nu_e \\ e^- \\ e^+ 
\end{array} \right)_L,
\left( \begin{array}{c}
\nu_\mu \\ \mu^- \\ \mu^+ 
\end{array} \right)_L,
\left( \begin{array}{c}
\nu_\tau \\\tau^- \\ \tau^+ 
\end{array} \right)_L
\sim (1,3,0)
\end{equation}
The first two families of quarks transform
as $SU(3)_L$ anti-triplets and the
third as a triplet.
\begin{equation}
Q_{1,2L}=
\left( \begin{array}{c}
d \\ u \\ D 
\end{array} \right)_L,
\left( \begin{array}{c}
s \\ c \\ S 
\end{array} \right)_L
\sim (3,\bar{3},-\frac{1}{3})
\end{equation}
\begin{equation}
Q_{3L}=
\left( \begin{array}{c}
t \\ b\\ T 
\end{array} \right)_L
\sim (3,3,+\frac{2}{3})
\end{equation}
All right-handed quarks transform as singlets.
Note that $D$, $S$ and $T$ are
exotic quarks, with charges
$-\frac{4}{3}$, $-\frac{4}{3}$ and
$\frac{5}{3}$ respectively.
The above fermion representations are
anomaly-free once all three generations
are included.

\subsection{Scalar Content} 
The minimal scalar content required 
to break symmetry and give all fermions a
realistic mass
consists of three triplets and one sextet:
\begin{equation}
\eta= \left( \begin{array}{c}
\eta^0 \\ \eta_1^- \\ \eta_2^+ 
\end{array} \right) 
\sim (3,0),
\rho= \left( \begin{array}{c}
\rho^+ \\ \rho^0 \\ \rho^{++}
\end{array} \right)
\sim(3,1),
\chi= \left( \begin{array}{c}
\chi^- \\ \chi^{--} \\ \chi^0 
\end{array} \right)
\sim (3,-1)
\end{equation}
\begin{equation}
S= \left( \begin{array}{ccc}
\sigma_1^0 & s_2^- & s_1^+ \\
s_2^- & S_1^{--} & \sigma_2^0 \\
s_1^+ & \sigma_2^0 & S_2^{++}
\end{array} \right)
\sim (6,0)
\end{equation}
These scalars have Yukawa couplings to the
quarks as follows:
\begin{eqnarray}
{\cal L}^{Yukawa}_{quarks} & = & 
h_u^{ik} \overline{Q}_{iL} u_{kR} \rho^\star 
+ h_d^{ik} \overline{Q}_{iL} d_{kR} \eta^\star 
+ h_D^{ij} \overline{Q}_{iL} D_{jR} \chi^\star
\\ \nonumber & &
+ h_d^{3k} \overline{Q}_{3L} d_{kR} \rho 
+ h_u^{3k} \overline{Q}_{3L} u_{kR} \eta 
+ h_T      \overline{Q}_{3L} T_{R} \chi
\end{eqnarray}
with $i,j=1,2$ and $k,l=1,2,3$,
while the Yukawa couplings of the leptons are given by
\begin{equation}
{\cal L}^{Yukawa}_{leptons}=
h^{kl}_A \bar{f}_{kL}f^{c}_{lR} \eta^\star +
h^{kl}_S \bar{f}_{kL} S f^{c}_{lR} 
\label{eq:Yukleptons}
\end{equation}
Written with explicit $SU(3)$ indices, these terms are
of the form
$\epsilon_{\alpha \beta \gamma} \bar{f}^\alpha f^{c\beta} \eta^{\star \gamma}$
and
$\bar{f}^\alpha S_{\alpha \beta} f^{c \beta}$
.
The scalar multiplets gain VEVs as follows:
\begin{equation}
\langle \eta \rangle = \left( \begin{array}{c}
v_1 \\ 0 \\ 0 
\end{array} \right) 
,
\langle \rho \rangle = \left( \begin{array}{c}
0 \\ v_2 \\ 0 
\end{array} \right)
,
\langle \chi \rangle = \left( \begin{array}{c}
0 \\ 0 \\ v_3 
\end{array} \right)
\label{eq:VEVS1}
\end{equation}
and 
\begin{equation}
\langle S \rangle = \left( \begin{array}{ccc}
v_4^\prime &  0 & 0  \\
0  & 0  & v_4 \\
0  & v_4 & 0  
\end{array} \right)
\label{eq:VEVS2}
\end{equation}
The symmetry breaking scheme
can be represented as:
\begin{equation}
SU(3)_L \times U(1)_X
\stackrel{\langle \chi \rangle}{\rightarrow}
SU(2)_L \times U(1)_Y
\stackrel{\langle \eta,\rho,S \rangle}{\rightarrow}
U(1)_Q
\label{eq:symbreak}
\end{equation}
That is, the triplet $\chi$ is responsible for breaking
$SU(3)_L \times U(1)_X$ symmetry, 
and therefore would receive a VEV
of greater value than $\rho$, $\eta$ and $S$, whose
VEVs would be around the electroweak breaking
scale. 
The exotic quarks and gauge bosons gain mass from $v_3$,
the Standard Model quarks gain mass from $v_1$ and $v_2$, while
the charged leptons gain mass from $v_1$ and $v_4$.
In the case of $v_4^\prime \neq 0$, neutrinos
develop a tree-level majorana mass.

\section{Global Symmetries}
The Yukawa couplings written above possess three
independent global symmetries which are not broken
by the VEVs $v_1$,$v_2$,$v_3$ and $v_4$.
As well as the $X$ charge already given we can
assign conserved charges
 ${\cal B}$ and ${\cal L}$ 
to the fermion and scalar mutiplets as follows:
\begin{eqnarray}
{\cal B} (f_{1,2,3L}) & = & 0 \\
{\cal B} (Q_{1,2L})   =  
{\cal B} (Q_{3L}) & = & \frac{1}{3} \\ 
{\cal B} (u_{1,2,3R}) =
{\cal B} (d_{1,2,3R}) & = & \frac{1}{3} \\ 
{\cal B} (D_{1,2R}) = 
{\cal B} (T_{R}) & = &
\frac{1}{3} \\
{\cal B} (\eta)  = 
{\cal B} (\rho) =
{\cal B} (\chi)  = 
{\cal B} (S) & = & 0
\end{eqnarray}
and
\begin{eqnarray}
{\cal L} (f_{1,2,3L}) & = & + \frac{1}{3} \\
{\cal L} (Q_{1,2L})   & = & + \frac{2}{3} \\ 
{\cal L} (Q_{3L}) & = & -\frac{2}{3} \\
{\cal L} (u_{1,2,3R}) =
{\cal L} (d_{1,2,3R}) & = & 0 \\ 
{\cal L} (D_{1,2R}) & = & + 2 \\ 
{\cal L} (T_{R}) & = &
-2  \\
{\cal L} (\eta)  = 
{\cal L} (\rho) & = & - \frac{2}{3} \\
{\cal L} (\chi) & = & + \frac{4}{3} \\
{\cal L} (S) & = & + \frac{2}{3}
\end{eqnarray}
The lepton and baryon number of the individual
components of each multiplet will then be given by:
\begin{eqnarray}
L & = & \frac{4}{\sqrt{3}} \frac{\lambda_8}{2} + {\cal L}I \\
B & = & {\cal B} I
\end{eqnarray}

\section{The Scalar Potential}
The full gauge-invariant scale potential was
first given in Ref \cite{A57}, and may be
written as
\begin{equation}
V=V_{LNC} + V_{LNV}
\end{equation}
where
\begin{eqnarray}
V_{LNC} & = & \mu_1^2 \etad + \mu_2^2 \rhod + 
\mu_3^2 \chid 
\\ \nonumber
 & + & 
 \lambda_1 (\etad)^2 + \lambda_2 (\rhod)^2
+ 
\lambda_3 (\chid)^2
\\ \nonumber
& + &   \lambda_4 (\etad)(\rhod) + \lambda_5
(\etad)(\chid) + \lambda_6 (\rhod)(\chid) \\ \nonumber
& + &
\lambda_7 (\rhodg \eta)(\etadg \rho) + \lambda_8 (\chidg \eta)
(\etadg \chi) + \lambda_9 (\rhodg \chi)(\chidg \rho)
\\ \nonumber
 & + & 
f_1 \left( \eta \rho \chi + H.c. \right)
\\ \nonumber
& + & 
\mu_4^2 Tr(\sds) + \lambda_{10} \left[ Tr(\sds) \right] ^2 + 
\lambda_{11}Tr \left[ (\sds)^2 \right] \\ \nonumber
& + &
\left[ \lambda_{12} (\etad) + \lambda_{13} (\rhod) + \lambda_{14}
(\chid) \right] Tr (\sds)
\\ \nonumber
& + & f_2 (\rho \chi S^\dagger + H.c.) \\ \nonumber
 & + & \lambda_{15} \eta^\dagger S S^\dagger \eta 
+ \lambda_{16} \rho^\dagger S S^\dagger \rho
+ \lambda_{17} \chi^\dagger S S^\dagger \chi \\ \nonumber
 & + & \lambda_{19} \rho^\dagger S \rho \eta 
+ \lambda_{20} \chi^\dagger S \chi \eta
+ \lambda_{21} \eta \eta SS + H.c.\\ \nonumber
\end{eqnarray}
and
\begin{eqnarray}
V_{LNV}
 & = & 
\bar{f}_3 \eta S^\dagger \eta +
\bar{f}_4 SSS +
\bar{\lambda}_{22} \chi^\dagger \eta \rho^\dagger \eta + \\ \nonumber
& &
\bar{\lambda}_{23} \eta^\dagger S \chi \rho +
\bar{\lambda}_{24} \chi \rho SS + H.c. 
\end{eqnarray}
The coefficients $f_1$,$f_2$, $\bar{f}_3$ and $\bar{f}_4$ have dimensions of mass,
and
bars have been used to denote terms which do not
conserve lepton number, ${\cal L}$, as defined above.

In previous studies, \cite{B64,331scalar}
the lepton number violating
terms have often been excluded, commonly by
the adoption of an appropriate discrete symmetry.
While there is no reason within the 331 Model
why such terms should not be present, 
experimental limits on processes which
do not conserve total lepton number
such as neutrinoless double beta decay,
\cite{NL105}
require them to be very small.

If the potential is restricted to $V_{LNC}$,
then two solutions exist which minimise the
potential and satisfy equations 
\ref{eq:VEVS1}
and \ref{eq:VEVS2}
above,
namely $v_4^\prime=0$ or 

\begin{eqnarray}
2 \lambda_{11}v_4^{\prime 2} & = & 
-\left( \lambda_{15} + 2\lambda_{21} \right)v_1^2 + \frac{\lambda_{16}}{2}v_2^2 
+ \frac{\lambda_{17}}{2}v_3^2 + \lambda_{11} v_4^2 
\\ \nonumber & - & 
 \frac{\lambda_{19}} {\sqrt{2}}\frac{v_1v_2^2}{v_4}
+ \frac{\lambda_{20}} {\sqrt{2}}\frac{v_1v_3^2}{v_4}
+ \frac{f_2} {\sqrt{2}}\frac{v_2v_3}{v_4}
\end{eqnarray}
The latter solution leads to the formation of
majorons (since lepton number is now
being broken spontaneously) and may be excluded by LEP.
\cite{NL91}

If the full potential is used, on the other hand,
a VEV for $v_4^\prime$ is automatically
induced \cite{A66} to satisfy the constraint
\begin{equation}
A v_4^{ \prime 3} 
+ \bar{B} v_4^{ \prime 2} 
+ C v_4^{ \prime} 
+ \bar{D} =0
\end{equation}
where
\begin{eqnarray}
A & = &
2 \lambda_{11} 
\\
\bar{B} & = & 
6\bar{f}_4 + \sqrt{2} \bar{\lambda}_{24} \frac{v_2v_3}{v_4}
\\
C & = & 
\left(
\lambda_{15} + 2 \lambda_{21} \right) v_1^2 - 
\frac{1}{2} \left( \lambda_{16}v_2^2 + \lambda_{17}v_3^2 \right)
- \lambda_{11} v_4^2
 \nonumber 
\\ & & 
+ \frac{1}{\sqrt{2}} \frac{v_1}{v_4}
\left(
\lambda_{19}v_2^2 -f_2 \frac{v_2 v_3}{v_1} - \lambda_{20}v_3^2
\right)
\\
\bar{D} & = & 
\bar{f_3} v_1^2 - 3\bar{f_4} v_4^2 - \bar{\lambda}_{23} v_1v_2v_3 - \sqrt{2}\bar{\lambda}_{24} v_2v_3v_4
\end{eqnarray}
that is,
\begin{equation}
v_4^\prime = \frac{-\bar{D}}
{C + \bar{B} v_4^\prime  + A v_4^{\prime 2}}
\end{equation}
In other words, 
$\langle \sigma_1^0 \rangle \simeq -\bar{D}/C$.
From equation \ref{eq:Yukleptons},
$m_{\nu_\tau} / m_\tau \sim v_4^\prime /v_4$
and assuming $v_4 \sim$ 100 GeV, to
obtain $m_{\nu_\tau} \simeq$ 0.1 eV 
requires $\bar{D}/ C \simeq 6 \times 10^{-9}$ GeV.
Note that this ratio is consistent with
other experimental constraints.
Apart from neutrinoless double beta decay mentioned
earlier, $v_4^\prime$ is also required to be small
to maintain a value for the $\rho$-parameter
sufficiently close to unity. \cite{NL89}

It is perhaps worth pointing out that, although
this ratio might seem large,
ratios of large order already
exist in the 331 Model in the Yukawa sector.
As another comparison,
in the Standard Model, the ratio of the top quark
mass to the electron mass of $3.4 \times 10^5$ is
due entirely to the relative strength of the respective
Yukawa couplings.

\section{Conclusion}

We have shown that a mechanism for generating
very small tree-level neutrino masses can easily
be implemented in the 331 Model. 

Lepton number violation occurs very naturally in
these models since the charged lepton of each
family and its antiparticle are placed in the
same triplet, and with
the one
assumption that lepton-number violating terms in
the scalar potential are sufficiently small,
 tree-level majorana neutrino masses 
of order 0.1 eV are automatically
induced without the introduction of any 
additional fermions or
scalar multiplets to the model.
We stress that no mass scale greater than 
a few TeV needs to be made use of, and
thus this model remains testable at the
next generation of colliders.

We also note briefly that, as
an alternative, a similar mechanism can also be 
employed in a second variant of the 331 Model, in which
a heavy lepton is included in the lepton triplet.
The leptons of each family then transform as follows:
\begin{equation}
\left(
\begin{array}{c}
l^- \\
\nu_l \\
L^+
\end{array}
\right)_L
\sim (1,3,0), 
l^-_R \sim(1,1,-1),
L^+_R \sim(1,1,+1)
\end{equation}
where $l^-$ is the conventional charged lepton of each
family and $L^+$ represents an exotic heavy lepton, 
both of which now obtain Dirac masses  
from the triplets $\rho$ and $\chi$ respectively. 

An interesting consequence  of this version of the model is that 
bileptons only
couple standard to exotic leptons 
and thus
many processes
which have been studied in order to place lower limits on
bilepton mass will not occur.  

In this scenario, the sextet can be made very heavy with only  
very small values for $v_4$ and $v_4^\prime$ induced.
However, since a new mass scale  $\mu_4 \sim 10^{12}$ GeV
must be introduced, we do not find this method as
attractive as the main one outlined.

\section*{Acknowledgements}
The authors would like to thank Dr Robert Foot
for helpful discussions.


\begin{thebibliography}{99}
\bibitem{SuperK}Super-Kamiokande Collaboration, 
Y.Fukada et al., Phys. Rev. Lett. {\bf 81}, 1562 (1998).
\bibitem{Ma1}E.Ma, Phys. Rev. Lett. {\bf 81}, 1171 (1998).
\bibitem{Zee}A.Zee, Phys. Lett. {\bf B93}, 389 (1980).
\bibitem{Seesaw}M.Gell-Mann, P.Ramond and R.Slansky, in
{\it Supergravity}, edited by P. van Nieuwenhuizen and D.Z.
Freedman (North-Holland, Amsterdam, 1979); 
T.Yanagida, in {\it Proceedings of the Workshop
on the Unified Theory and the Baryon Number in the Universe},
edited by O.Sawada and A.Sugamoto (KEK Report No. 79-18,
Tsukuba, Japan, 1979);
R.N.Mohapatra and G.Senjanovi{\'c}, Phys. Rev. Lett. {\bf 44},
912 (1980).
\bibitem{Ma2}R.N.Mohapatra and G.Senjanovi{\'c}, Phys. Rev. {\bf D23},
165 (1981); 
E.Ma and U.Sarkar, Phys. Rev. Lett. {\bf 80}, 5716 (1998).
\bibitem{Ma3}E.Ma, Phys. Rev. Lett {\bf 86}, 2502 (2001).
\bibitem{Valle}J.W.Valle and M.Singer, Phys. Rev. {\bf D28}, 540 (1983).
\bibitem{B0}F.Pisano and V.Pleitez, Phys. Rev. {\bf D46}, 
410 (1992).
\bibitem{A0}P.H.Frampton, Phys. Rev. Lett. {\bf 69}, 
2889 (1992).
\bibitem{B64}R.Foot, O.F.Hern\'andez, F.Pisano and V.Pleitez, 
Phys. Rev. {\bf D47}, 4158
(1993).
\bibitem{A75}D.Ng, Phys. Rev. {\bf D49}, 4805 (1994);
J.T.Liu, Phys. Rev. {\bf D50}, 542 (1994).
\bibitem{A57}J.T.Liu and D.Ng, Phys. Rev. {\bf D50}, 548 (1994).
\bibitem{Yasue}Radiative mechanisms for generating neutrino mass have
also been recently considered in different variants of the 331 Model: 
Y.Okamoto and M.Yasue, Phys. Lett. {\bf B466}, 267 (1999); 
T.Kitabayashi and M.Yasue, hep-ph/0006040;
T.Kitabayashi and M.Yasue, Phys. Rev. {\bf D63}, 095002 (2001).
\bibitem{other331}Other recent work includes:
P.Das and P.Jain, Phys. Rev. {\bf D62}, 075001 (2000);
V.Pleitez, Phys. Rev. {\bf D61}, 057903 (2000);
M.B.Tully and G.C.Joshi, Phys.  Lett. {\bf B466}, 333 (1999); 
Y.A.Coutinho, P.P.Queiroz Filho and M.D.Tonasse, Phys. Rev. {\bf D60}, 115001 (1999);
H.N.Long and  V.T.Van, J.Phys. {\bf G25}, 2319 (1999);
G.Taveres-Velasco and J.J.Toscano, Europhys. Lett. {\bf 53}, 465 (2001);
H.N.Long, Yad. Fiz. {\bf 64}, 106 (2001) [ Phys. At. Nucl. {\bf 64}, 103 (2001) ];
P.H.Frampton and A.Rasin, Phys. Lett. {\bf B482}, 129 (2000);
N.A.Ky, H.N.Long and D.V.Soa, Phys. Lett. {\bf B486}, 140 (2000);
A.Doff and F.Pisano, Mod. Phys. Lett. {\bf A15}, 1471 (2000);
J.E.Cieza Montalvo and M.D.Tonasse, hep-ph/0008196;
H.N.Long and L.P.Trung, Phys. Lett. {\bf B502}, 63 (2001).
\bibitem{331scalar}
M.D.Tonasse, Phys. Lett. {\bf B381}, 191 (1996); 
N.T.Anh, N.A.Ky and H.N.Long, Int. J. Mod. Phys. {\bf 15}, 283 (2000);
M.B. Tully and G.C.Joshi, hep-ph/9810282.
\bibitem{NL105}J.C.Montero, C.A.de S.Pires and V.Pleitez, hep-ph/0003284.
\bibitem{NL91}J.C.Montero, C.A.de S.Pires and V.Pleitez, Phys. Rev.
{\bf D60}, 115003 (1999).
\bibitem{A66}P.H.Frampton, P.I.Krastev and J.T.Liu, Mod. Phys. Lett. {\bf A9}, 761 (1994).
\bibitem{NL89}J.C.Montero, C.A.de S.Pires and V.Pleitez, Phys. Rev.
{\bf D60}, 098701, (1999).
\end{thebibliography}
\end{document}